\pdfoutput=1
\documentclass[11pt, a4paper]{article}
\usepackage{ifpdf}
\usepackage{hyperref}
\hypersetup{pdfstartview=FitH, colorlinks=true, urlcolor=blue, citecolor=blue, linkcolor=blue}
\usepackage[margin=1in]{geometry}
\usepackage{url}
\usepackage[utf8]{inputenc}
\usepackage[T1]{fontenc}
\usepackage[english]{babel}
\usepackage{times}
\usepackage{amsfonts}
\usepackage{authblk}
\usepackage{booktabs}
\usepackage{enumitem} 
\usepackage{tikz}
\usepackage{float}
\usepackage{array}
\usepackage{footnote}

\usepackage{color}

\usepackage{tabularx}
\usepackage{longtable}
\usepackage{graphicx,multirow}
\usepackage{multicol}

\usepackage{makecell}
\usepackage{ctable} 
\usepackage{pifont} 
\usepackage{xspace} 
\usepackage{algorithm}
\usepackage{algpseudocode}
\usepackage{amsmath}
\usepackage{amssymb}
\usepackage{mathtools}
\usepackage{amsthm} 
\usepackage{ragged2e} 
\usepackage[none]{hyphenat} 

\usepackage{caption}

\newcolumntype{C}[1]{>{\centering\arraybackslash\hspace{0pt}}m{#1}}
\newcolumntype{Y}{>{\centering\arraybackslash}X}

\floatname{algorithm}{Function}

\newcommand\rr[1]{\textcolor{blue}{~#1}}

\newcommand{\scheme}{BDIMHS\xspace}
\newcommand{\trustee}{\textit{Trustee}\xspace}
\newcommand{\stewards}{\textit{Stewards}\xspace}
\newcommand{\trustanchor}{\textit{Trust Anchor}\xspace}
\newcommand{\issuer}{\textit{Issuer}\xspace}
\newcommand{\holder}{\textit{Holder}\xspace}
\newcommand{\verifier}{\textit{Verifier}\xspace}
\newcommand{\steward}{\textit{Steward}\xspace}
\newcommand{\pt}{\textit{Patient}\xspace}
\newcommand{\hp}{\textit{Healthcare personnel}\xspace}
\newcommand{\hosp}{\textit{Hospital}\xspace}
\newcommand{\gov}{\textit{Government}\xspace}
\newcommand{\nphs}{\textit{NPHS}\xspace}
\newcommand{\cl}{\textit{Clinic}\xspace}
\newcommand{\rl}{{\textit{Research laboratory}}\xspace}
\newcommand{\authapp}{{\textit{Authorization application}}\xspace}

\definecolor{lime}{HTML}{A6CE39}
\DeclareRobustCommand{\orcidicon}{
	\begin{tikzpicture}
	\draw[lime, fill=lime] (0,0) 
	circle [radius=0.16] 
	node[white] {{\fontfamily{qag}\selectfont \tiny ID}};
	\draw[white, fill=white] (-0.0625,0.095) 
	circle [radius=0.007];
	\end{tikzpicture}
	\hspace{-2mm}
}

\foreach \x in {A, ..., Z}{
	\expandafter\xdef\csname orcid\x\endcsname{\noexpand\href{https://orcid.org/\csname orcidauthor\x\endcsname}{\noexpand\orcidicon}}
}

\title{Blockchain-based Decentralized Identity Management for Healthcare Systems}

\author{Arnaf Aziz Torongo\orcidA{} \hspace{3.2cm} Mohsen Toorani\orcidB{}\\
\hspace{0.5cm}\texttt{250797@usn.no} \hspace{2.2cm} \texttt{mohsen.toorani@usn.no} \\ 
\vspace{10pt}
Department of Science and Industry Systems \\ University of South-Eastern Norway \\  Kongsberg, Norway \vspace{-30pt}} 
\date{}

\begin{document}
\maketitle
\let\thefootnote\relax\footnotetext{\scriptsize{Copyright  \copyright ~Authors 2023. All Rights Reserved. This is the author's version of the work. It is posted here for your personal use, not for redistribution. A variant will appear in Proceedings of 18th IFIP Summer School on Privacy and Identity Management 2023}.}

\begin{abstract}
Blockchain-based decentralized identity management provides a promising solution to improve the security and privacy of healthcare systems and make them scalable. Traditional Identity Management Systems are centralized, which makes them single-point-of-failure, vulnerable to attacks and data breaches, and non-scalable. In contrast, decentralized identity management based on the blockchain can ensure secure and transparent access to patient data while preserving privacy. This approach enables patients to control their personal health data while granting permission for medical personnel to access specific information as needed. We propose a decentralized identity management system for healthcare systems named BDIMHS based on a permissioned blockchain with Hyperledger Indy and Hyperledger Aries. We develop further descriptions of required functionalities and provide high-level procedures for network initialization, enrollment, registration, issuance, verification and revocation functionalities. The proposed solution improves data security, privacy, immutability, interoperability, and patient autonomy by using selective disclosure, zero-knowledge proofs, Decentralized Identifiers, and Verifiable Credentials. Furthermore, we discuss the potential challenges associated with implementing this technology in healthcare and evaluate the performance and security of the proposed solution. 
\end{abstract}

\section{Introduction}\label{sec:introduction}
Identity Management (IDM) is essential for the security and privacy of modern systems. The modern healthcare system is a vast domain, and it may contain numerous amounts of patient data in various formats, which are extremely sensitive \cite{BourasLZWZN20}. To get the most out of this data, a complex collaboration is always required among the administrators, physicians, patients and other healthcare personnel. As a result, identity management of this system is equally essential as it possesses all the private medical information and records. Most healthcare systems use traditional Identity Management Systems (IDMS) that are centralized and managed by central authorities that act as Identity Providers \cite{XiangCF22}. Thus, identity owners have no control over their data and cannot prevent any exploitation of their own identities and privacy \cite{JavedABMCQ21}. Most importantly, there exists a challenge to achieving interoperability among different services. 
Blockchain-based decentralized identity, broadly known as Self-Sovereign Identity (SSI), can lay the foundation of trust by giving ownership and control of identity to the users \cite{Cucko22}. \issuer, \textit{Identity Owner} or \holder, and \verifier are the three core components of the decentralized ecosystem that create the trust triangle for the decentralized system. Decentralized Identifier (DID) and Verifiable Credential (VC) are the two pillars of decentralized identity that create the foundation of distributed ledger-based verifiable digital identity, entirely governed by its owner \cite{NaikJ21}. There is a pivotal relationship between the electronic healthcare system and IDMS as healthcare data requires a high level of security and privacy. 

Maintaining access controls, authentication, privacy, auditability, integrity, and non-repudiation of records are the major security requirements affiliated with the healthcare systems. Apart from security, there are functionality and usability requirements comprising scalability, interoperability and ease of use when it comes to managing the identities of the system. The users may be reluctant to disclose the sensitive information required for healthcare due to the risk of identity theft or misuse. Decentralized Identity Management (DIM) can eliminate the intermediaries involved in traditional IDMS. As end-user control is de-emphasized in a traditional IDMS, our motivation is to investigate and implement the potentiality of blockchain-based DIM and its possible implementation in healthcare systems. 

Decentralized identity management has emerged as a highly researched and promising field, offering numerous benefits for healthcare applications. 
Bouras et al. \cite{BourasLZWZN20} discussed decentralized identity management in electronic health systems, focusing on protecting personal data. They compared existing solutions based on identity management principles.
Tanwar et al. \cite{TanwarPE20} proposed a symmetric key cryptography-based access control policy algorithm for healthcare data interoperability using Hyperledger. They evaluated performance metrics and reviewed blockchain-based EHR systems.
Javed et al. \cite{JavedABMCQ21} proposed a blockchain-based decentralized identity management system with unique health identifiers. They implemented a smart contract on Ethereum and measured performance metrics.
Madine et al. \cite{MadineSJYAEC20} proposed blockchain-based Personal Health Record (PHR) architecture with multi-party authorization and cryptographic threshold schemes. They used off-chain storage and evaluated security and limitations.
Xiang et al. \cite{XiangWF20} presented a permissioned blockchain-based identity management and user authentication scheme for e-health, resilient to various attacks and satisfying medical security requirements.
Mikula et al. \cite{MikulaJ18} demonstrated a prototype of blockchain-based identity and access management for EHR using Hyperledger Fabric. They emphasized the use of private or consortium blockchain for healthcare institutions.
Jiang et al. \cite{JiangCWYMH18} proposed BlocHIE, a blockchain-based medical data exchange platform using EMR-Chain and PHD-Chain for preserving healthcare data. They combined off-chain storage and on-chain verification to address privacy and authenticity requirements.
Manoj et al. \cite{ManojMN2022} proposed a blockchain-based framework for patient authentication and consent management using Hyperledger Indy blockchain and Aries. They demonstrated the generation and verification of verifiable credentials for EHR access.
Saidi et al. \cite{SaidiLAME22} proposed a Decentralized Self-Management of Data Access Control (DSMAC) using DID-supported smart contracts and Role-based access control. They focused on the security requirements of SSI in the healthcare system.

In this paper, we propose a DIM that creates the missing trust layer across various healthcare platforms and provides sustainability, reliability, availability, and security by abolishing intermediaries and empowering users to control their own healthcare identities. We present the design of the \scheme scheme for healthcare systems. We provide a Proof of Concept (PoC) implementation, conduct a heuristic security analysis, and compare the security and privacy goals with other proposed solutions. Additionally, a detailed performance evaluation is conducted to assess the feasibility and effectiveness of the proposed scheme in healthcare systems.

The rest of this paper is organized as follows: Section \ref{sec:dim} provides an overview of the theoretical foundations of DIM and related terminology. Section \ref{sec:proposed_scheme} introduces the \scheme scheme. In Section \ref{sec:security_analysis}, a security analysis is conducted, including threat modeling, countermeasures, and security requirements. The performance evaluation is discussed in Section \ref{sec:performance_analysis}. Finally, Section \ref{sec:conclusion} concludes the paper.

\section{Decentralized Identity Management}\label{sec:dim}
A decentralized identity is an alternative approach compared to traditional identity management, which enables individuals or organizations to fully govern their digital identities using distributed technology under the hood. It influences individuals to have full possession and control over their identity information, trimming the dependency on centralized entities which store, manage and use personal data in their domain. As opposed to this, individuals can store their identity attributes on their devices or in identity wallets. They can decide when, whom, and how to share the credentials with the Service Providers (SP). In this section we will describe briefly the key elements of DIM.

\subsection{Decentralized Identifier}
DIDs are the foundational building blocks of decentralized identity. DIDs are unique identifier that delegates verifiable, decentralized digital identity by citing individuals, organizations to objects or any entity as determined by the controller of the DID \cite{W3C}. A DID is a short unique string consisting of three different parts. The first part is DID URI scheme identifier, the second part is DID method that identifies the ledger where the DID originates, and the last part is DID method-specific unique identifier \cite{W3C}. DIDs are designed to be resolvable to DID documents. 

\subsection{Verifiable Credentials}
VC is a W3C open standard model of credentials on the web in a way that is "cryptographically secure, privacy respecting, and machine-verifiable" \cite{W3CVC}. The VC can contain the same data a physical credential might have. Using cryptography like digital signature transforms VC more reliable and tamper-proof than physical credentials \cite{W3CVC}. With the VC model, developing applications with a solid digital basis of trust becomes possible. The cryptographic mechanism used in verifiable credentials ensures their integrity and authenticity. 

\subsection{Zero-knowledge Proof and Selective Disclosure}
A ZKP is a cryptographic technique for demonstrating that individuals have the required validity of the value of an attribute without disclosing the attribute's actual value. Prover can convince the verifier that a required statement is genuine by providing factual information about the witness \cite{YangL20}. ZKP solve the problem of identity theft and lack of privacy by eradicating the need to reveal personal information to prove the validity of claims. In decentralized identity verifiable presentations, a ZKP is used to demonstrate that the verifiable credentials were provided to the \holder without disclosing a specific, relatable identifier of the \holder to the \verifier. The \verifier can still obtain the requested claims, but the process of proving does not automatically disclose a unique identifier for the \holder.

One of the valuable privacy capabilities of VC is the support of selective disclosure. Selective disclosure allows proving a particular set of claims from the VC without revealing the entire attributes of the credentials. When users share information, they can select what information they want to share with the SPs. This method gives greater control to the user on how their data will be managed.

\subsection{Hyperledger Indy}
Hyperledger Indy is an open-source distributed ledger software that furnishes the layer for decentralized identity by providing a distributed ledger platform designed specifically for DIM. Indy is a public permissioned ledger. 
Indy provides various tools and reusable components like \textit{Indy Node}, \textit{Indy Plenum}, and \textit{Indy SDK}. Indy Node is responsible for running the nodes which maintain the state of the network and processes transactions. Each node holds its own distributed ledger and stores public records like public keys, service endpoints, credential schemas and credential definitions. Indy Plenum is an Redundant Byzantine Fault Tolerant (RBFT) based consensus protocol that Indy Node uses to reach consensus among the nodes of the network that provides a high level of fault tolerance.

Some pre-approved roles exist for some of the participants responsible for building trust in the Indy network \cite{TheodouliMVTLS20}. These two roles are \trustee and \stewards. 
\steward is the first actor of the Indy network and is responsible for onboarding other participants with \trustanchor role, also known as \textit{Endorser}. User with \textit{Endorser} role has the write permission for publishing DIDs, writing schemas and credential definitions on the ledger. The public DIDs of every \textit{Endorser} will be available in the ledger.

\begin{figure}[!t]
    \centering
    \includegraphics[width=0.85\textwidth]{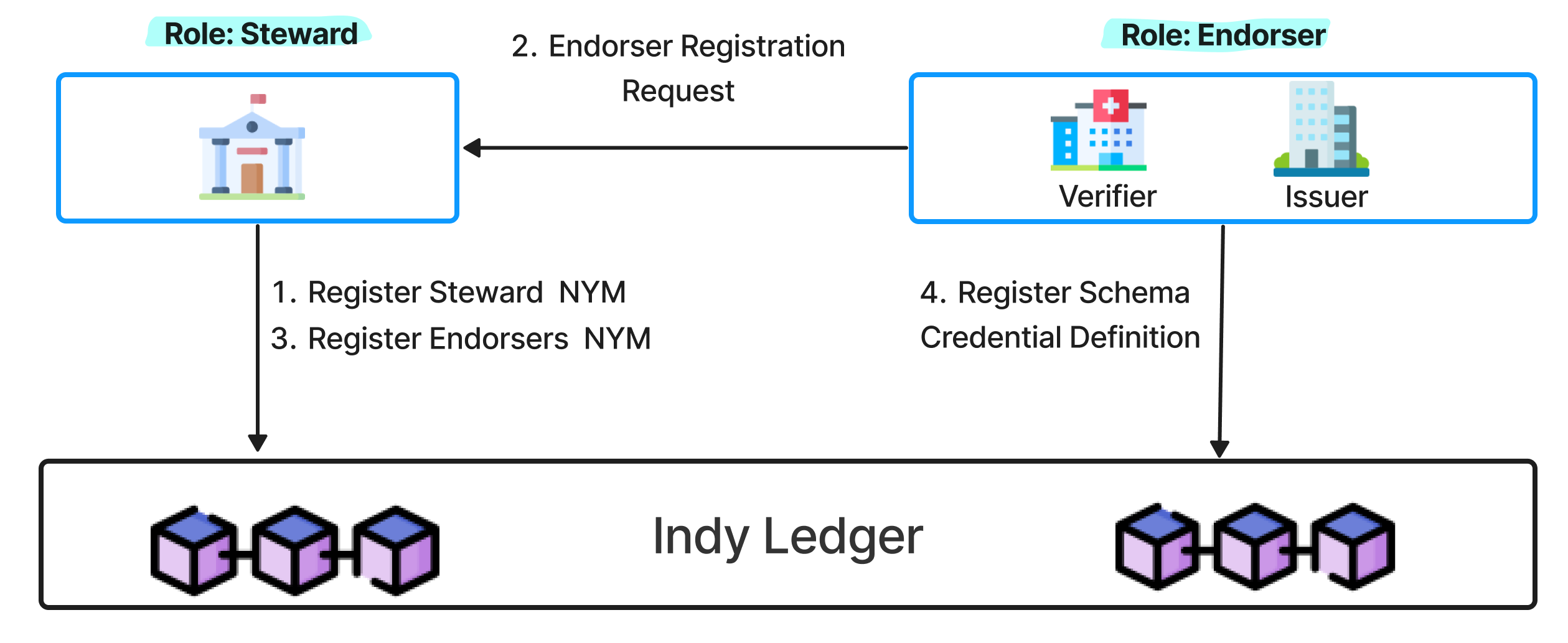} 
    \caption{Stewards and Endorsers Role}
    \label{fig:roles-indy}
\end{figure}

\subsubsection{Types of Indy Ledger}
There are multiple ledgers in Indy, more specifically on Indy Plenum. Audit Ledger, Pool Ledger, Config Ledger and Domain Ledger. 

\begin{itemize}
    \item \textbf{Pool Ledger: } The pool is generated from the genesis transactions. It stores the records of all the nodes added and removed. There are also the records of which nodes act as validator nodes and which are observers or participants of the Indy network. Each validator's unique identifier and public key are stored in the pool ledger. It also contains metadata about the validators, such as their IP address and other configuration specifications. Validators use a pool ledger to validate messages from other validators.
    
    \item \textbf{Audit Ledger: } Audit ledger maintains the synchronization between different types of the ledger by maintaining transaction order. It also initiates the pool ledger during startup. 

    \item \textbf{Config Ledger: } The Config ledger stores all the network-related configuration data, parameters, and transaction validation set by Governance Board. The changes to the network configuration are tracked here. 

    \item \textbf{Domain Ledger: } Domain ledger is heart of decentralized identity. It holds all the identity-related information and application-specific transactions.
    All the public records like Public DIDs, Credential Definition, Credential Schemas and Revocation info are stored in the Domain ledger.
\end{itemize}

    \begin{figure}[!t]
        \centering
        \includegraphics[width=0.75\textwidth]{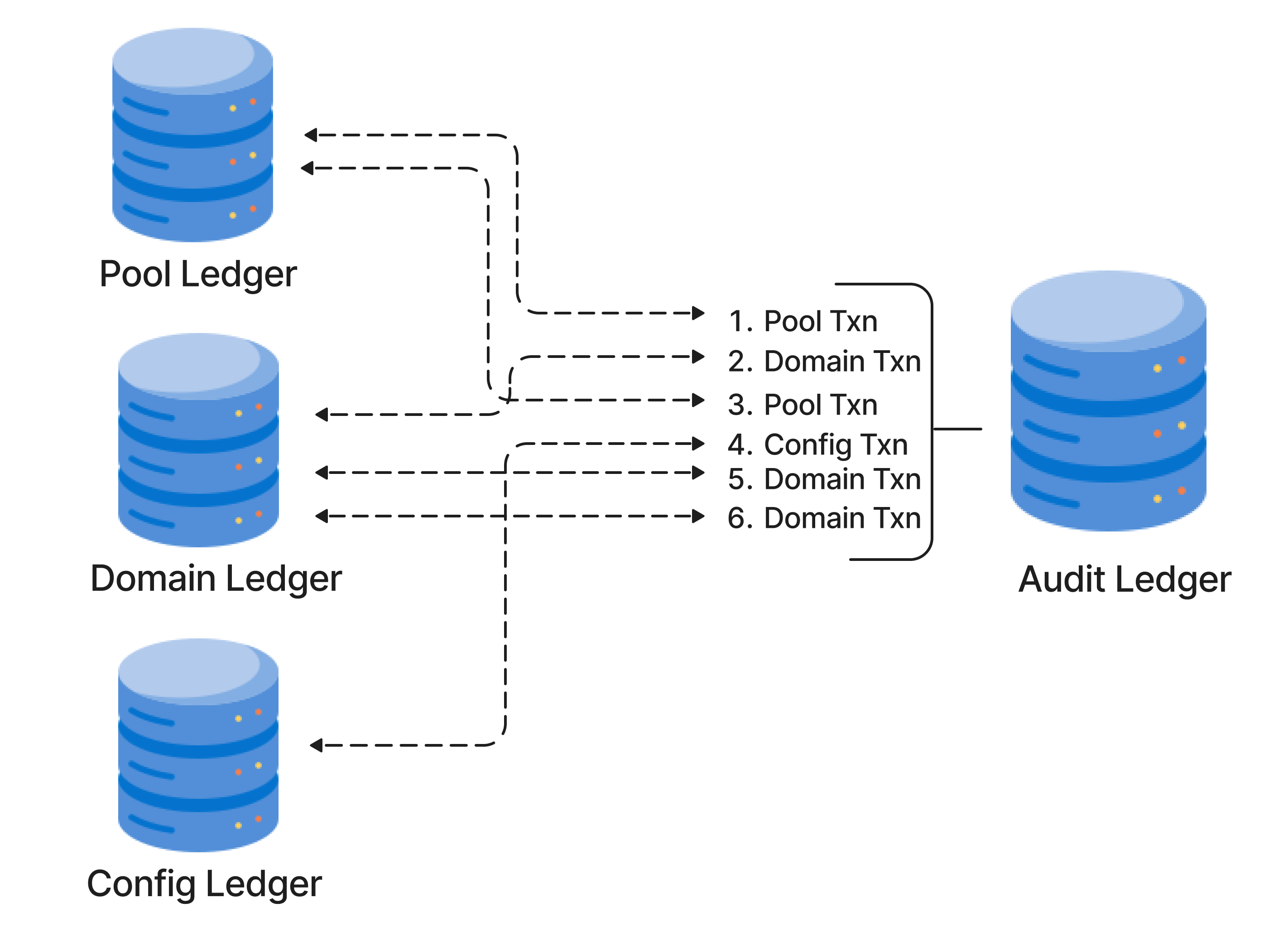}
        \vspace{-10pt}
        \caption{Audit Ledger}
        \label{fig:audit_ledger}
    \end{figure}

    \begin{figure}[!t]
        \centering
        \includegraphics[width=0.75\textwidth]{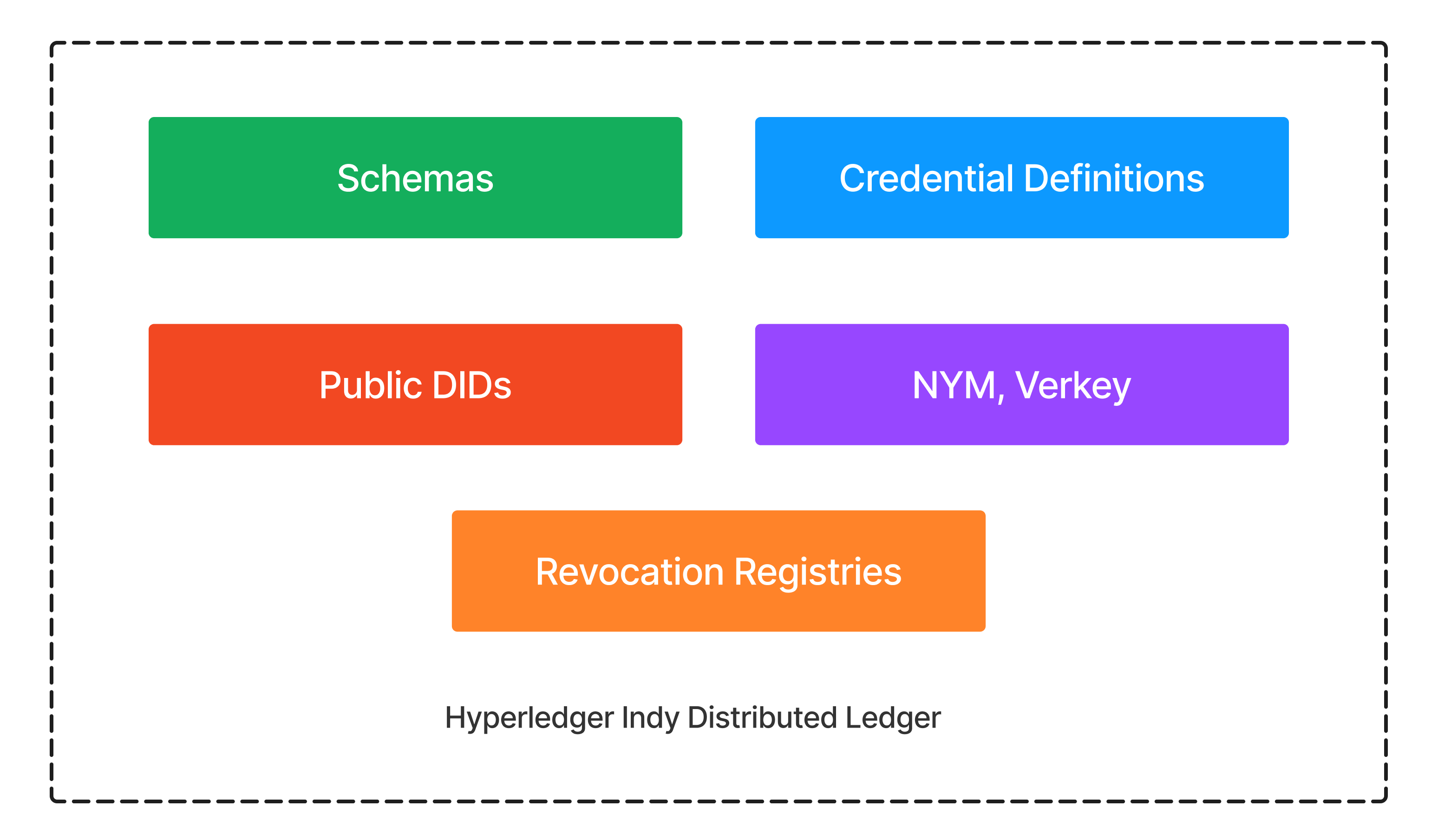} 
        \vspace{-10pt}
        \caption{Domain Ledger}
        \label{fig:dlt}
    \end{figure}

\subsection{Hyperledger Aries}
Hyperledger Aries is an open-source project that provides a set of tools and libraries for building secure and interoperable identity solutions on top of Hyperledger Indy. It can handle, store, and manage digital identities in a secure, decentralized way while ensuring privacy and giving privacy control to the users. 
Hyperledger Aries provides many APIs and SDKs with various platforms that make it easy to build and integrate business-specific identity solutions. Aries agents acts on behalf of a single identity owner, interacts using DIDComm and holds the cryptographic secret keys for its unique representation. It consists of two main components. One is the main framework, and another is the controller.
The controller is the main interface that drives the agent for receiving responses, requests from other agents, and initiating their requests. Each agent exposes a set of REST APIs and Webhooks that the controller can absorb. It holds all the business rules and conditions and acts based on the role defined in each verifiable credential.
Aries uses DIDComm for communication. DIDComm messaging provides a secure, private communication channel built on top of the decentralized design of DIDs \cite{didcomm}. 

\subsection{Terminologies of Hyperledger Indy and Aries}
 In this section, we will explore the key terminologies associated with Hyperledger Indy and Aries and give a brief description of them. 

\begin{itemize}
    \item \textbf{Schema Definition:}
    A Schema object is a machine-readable definition that defines a set of attributes and format \textit{Issuers} can use to issue VC within a Hyperledger Indy network \cite{AnonCreds}. It is a structure that represents all the attributes one credential can hold, like names and data types. 
    Schemas have a name and version and can be read from a Hyperledger Indy Node by any client to achieve standardisation among issuers \cite{AnonCreds}.
    
    \item \textbf{Credential Definition:}
    In Hyperledger Indy, VCs are created based on a Credential Definition known as CRED\_DEF. Credential Definitions reference the appropriate schema to use the specific attributes it will use. It contains a reference to the claim \issuer and its keys, a reference to the schema, and details about the issuance, and revocation criteria for the credentials \cite{SoltaniNA18}.

   \item \textbf{Revocation Registry:}
    Hyperledger Indy provides the functionality to revoke credentials once they have been issued to a \holder. Hypeledger Indy use Cryptographic Accumulators to handle credential revocation. The revocation registry contains information on whether the issuance credential has been revoked. The \verifier checks the revocation registry before verifying the credential on the ledger \cite{indyDID}.
    
    \item \textbf{Steward:}
    \steward can enrol new actors and assign \trustanchor role. All the organizations, like the \issuer, \verifier will be enrolled by \steward and assigned the \trustanchor or \textit{Endorser} role. \stewards establish the trust in the network, and the governing body's responsibility is to choose the proper Stewards. 

    \item \textbf{Trust Anchor/Endorser:} 
    The government or authority that is selected by the governing body to issue and verify digital credentials on the network have the role \trustanchor or \textit{Endorser}.

    \item \textbf{Decentralized Key Management Service:}
    Decentralized Key Management Service (DKMS) is a cryptographic key management procedure without relying on any central authority \cite{DKMSGithub}.
    It is a component within the Aries framework that provides a decentralized way to secure storage for securely managing and storing DIDs,  secret keys, and other information the Aries agent collects. It standardizes the agent wallet and offers support for the key recovery of the agents.

\end{itemize}

\section{Proposed Scheme (\scheme)}\label{sec:proposed_scheme}
We proposed a new decentralized user-centric passwordless identity management for the healthcare system, called \scheme. At first, we have identified the system's actors and core components. Then, we discuss the proposed scheme's functionalities, architecture, and workflow. In \scheme, a few actors have been considered out of countless actors related to healthcare. \pt, \hp, \gov, \textit{National Public Health Service} (NPHS), \hosp, \cl and \rl are the main actors in this scheme. As this system is user-centric, the whole system architecture has been designed primarily focusing on the \pt and \hp. In this proposed architecture, several key components have been identified that work together to support various functionalities. 

\begin{itemize}
    \item \textbf{Indy Node}: The Hyperledger Indy node manages the blockchain-based ledger. It would store the public DIDs, Schemas, Credential Definitions, NYM transactions, Revocation registries, and corresponding verification keys. As a public permissioned blockchain Indy is designed so that everyone on the network can read the blockchain's contents, but only \steward are permitted to participate in the validation procedure.
    
    \item \textbf{Agent}: Each user, such as the \pt or \hp has their own SSI Agent Wallet, which stores and manages their identities and correlated issued VCs, which contain their personal information. The agent is the software or controller that enables and manages peer-to-peer connections between other agents using private and pairwise DIDs.
    
    \item \textbf{Identity Provider}: The IdP would be responsible for issuing identities to the \holder. It issues associated credentials to the \textit{Holders} upon request and would also manage the revocation of identities if necessary. 
    
    \item \textbf{Authorization Application}: The \authapp acts as an authorization server. Each authorization server has its agent, which is connected to the ledger. All the authentication server is isolated and managed by their entities and connected with their system to perform Role-based access control (RBAC).
\end{itemize}

\subsection{Functionalities}
\scheme leverages the core functionalities developed and exposed through Hyperledger Indy and Hyperledger Aries to build a robust and secure identity management system. These functionalities, as described below, are presented in  Algorithms \ref{alg:init}--\ref{alg: revocation}. The corresponding notations used in the algorithms are explained in Table \ref{tab:notations}. 

\begin{figure}[!t]
    \centering
    \includegraphics[width=0.85\textwidth]{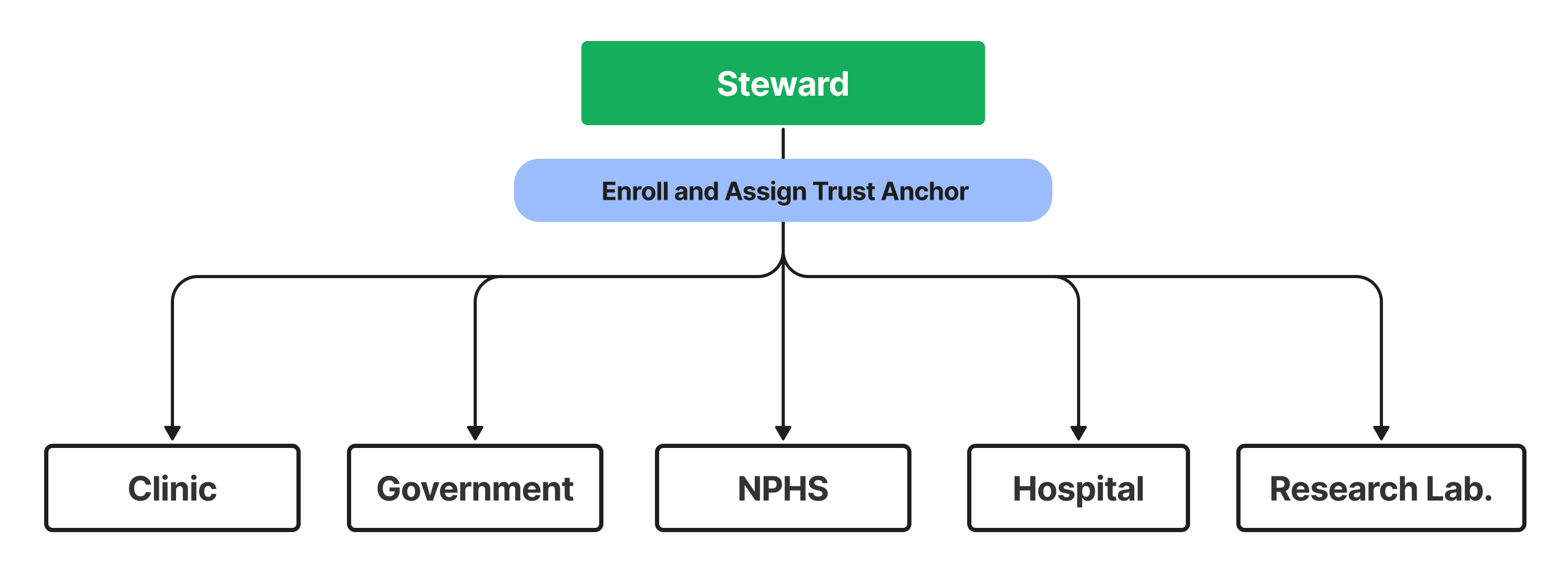} 
    \caption{Actor enrollment with Trust Anchor Role}
    \label{fig:onboardactor}
\end{figure}

\paragraph{\textbf{Initialization and Enrollment:}}
After connecting with the ledger using genesis transactions, \steward is responsible for creating and enrolling other actors in the network. A unique pairwise identification DID is created during the enrollment process between two parties. This unique DID creates DIDComm between two parties. Each DID contains a private signing key, a verifying key known as verkey, and a DID. Although this verkey and DID are recorded on the blockchain, private DIDs are not used to identify the actors from the ledger; instead used to create a communication channel. After onboarding, each actor creates a new DID, which consists of a signing key, verkey and a DID  known as \textit{Verinym} and then sends the Verinym to the \steward using the DIDComm protocol. Steward stores and grants the \trustanchor role while writing it in the ledger. Using the same procedure here, \steward enrols the actors' \gov, \nphs, \hosp, \cl and \rl depicts in Figure~\ref{fig:onboardactor}.

\paragraph{\textbf{Schema Registration:}}
After successful enrollment, the actors like \gov and \nphs can register Credential Schemas. \gov creates the NID schema, and NPSH creates the PID schema. Each schema contains the schema name, version, and necessary attributes stored on the ledger. These schemas are used to create specific Credential Definition . The Credential Definition holds the record of the schema it uses, the signature type, tag, revocation and information of the \issuer initiating the Credential Definition.

\paragraph{\textbf{Issuance:}}
The \ issuer only executes this functionality. The \gov and \nphs create a credential offer based on one specific Credential Definition and send the offer to the \pt and \hp. If the \holder accepts the offer, then the \issuer sends the issued VC.

\paragraph{\textbf{Verification:}}
This functionality is initiated from the \verifier side to the \holder. The \hosp, \cl or \rl create proof requests based on standard credential schema definition and business requirements. The proof request is sent to the \holder. After getting the response from the \holder, the \verifier verifies the proof and its non-revocation status presented by the \holder on the ledger to determine the validity.

\paragraph{\textbf{Revocation:}}
Revocation is the capability for an \issuer to publish that an issued VC is invalid. As the ledger is tamper-proof to update any credential change, revocation functionality becomes handy to revoke the existing credentials and issue a new one with updated attributes. The revocation list is maintained through a revocation registry and published on the ledger.

\begin{multicols}{2}
\begin{algorithm}[H]
\caption{Initialization}\label{alg:init}
\begin{algorithmic}
\State $ \mathsf{genTxnPath \leftarrow  GetPoolGenTxnPath(pool)} $
\State $ \mathsf{CreatePoolLedgerCnf(\{pool, genTxnPath\})} $
\State $ \mathsf{CreateWallet(\{stewardName, seed\})} $
\State $ \mathsf{SW \xleftarrow{W}  CreateAndStoreDID()} $
\end{algorithmic}
\end{algorithm}

\columnbreak

\begin{algorithm}[H]
\caption{Schema Registration}\label{alg: prepareCredential}
\begin{algorithmic}
\State $ \mathsf{schema \leftarrow \{name, version, attributes\}} $ 
\State $ \mathsf{req \leftarrow BuildSchemaReq(did, schema)} $
\State $ \mathsf{IL  \xleftarrow{W}  SignAndSubmitReqreq)} $
\State $ \mathsf{credDefReq \leftarrow BuildCredDefReq(did, schemaId)} $ 
\State $ \mathsf{IL \xleftarrow{W} SignAndSubmitReq(credDefReq)} $
\end{algorithmic}
\end{algorithm}
\end{multicols}

\begin{multicols}{2}
\begin{algorithm}[H]
\caption{Enrollment}\label{alg: enroll}
\begin{algorithmic}
\State $ \mathsf{req \leftarrow  \{did, nonce\}} $
\State $ \mathsf{ConnectReq(req)} $
    \If{$ \mathsf{response[nonce] = decrypt(req[nonce])}$}
            \State $\mathsf{return 1}$
    \EndIf

\State $ \mathsf{decrypted \leftarrow  decrypt(\{did, verkey, role\})} $   
\State $ \mathsf{nymReq \leftarrow BuildNymReq(did, decrypted)} $ 
\State $ \mathsf{IL \xleftarrow{W}  SignAndSubmitReq(nymReq)} $
\end{algorithmic}
\end{algorithm}

\columnbreak

\begin{algorithm}[H]
\caption{Verification}\label{alg: verification}
\begin{algorithmic}
\State $ \mathsf{req \leftarrow  \{connId, credDefId, requiredAttributes\}}$ 
\State $ \mathsf{response \leftarrow SendProofRequest(req)}$
\State $ \mathsf{IL \xleftarrow{R}  VerifyPresentation(response[presExId])}$ 
    \If{$ \mathsf{verified = true}$}
        \State $ \mathsf{return 1}$
    \Else
        \State $ \mathsf{return 0}$
    \EndIf
\end{algorithmic}
\end{algorithm}
\end{multicols}

\begin{multicols}{2}
\begin{algorithm}[H]
\caption{Issuance}\label{alg: issuance}
\begin{algorithmic}
\State $ \mathsf{req \leftarrow  \{did, schemaId, credDefId, attributes\}} $ 
\State $ \mathsf{credOffer \leftarrow CreateCredentialOffer(req)} $
\State $ \mathsf{response \leftarrow SendOffer(credOffer)}$ 
\State $ \mathsf{receiveAck \leftarrow IssueCred(response[credExId])}$ 
\end{algorithmic}
\end{algorithm}

\columnbreak

\begin{algorithm}[H]
\caption{Revocation}\label{alg: revocation}
\begin{algorithmic}
\State $ \mathsf{req \leftarrow  \{connId, credDefId, revRegId\}}$ 
\State $ \mathsf{response \leftarrow RevokeCredential(req) }$
\State $ \mathsf{IL \xleftarrow{W}  PublishRevocation(response[revRegId])}$
\State $ \mathsf{IL \xleftarrow{W}  ClearPendingRevocation()}$
\end{algorithmic}
\end{algorithm}
\end{multicols}

\begin{table}[H]
  \centering
  \footnotesize
  \caption{Description of Notations}
  \begin{tabularx}{0.88\textwidth}{|p{2.5cm}|p{3.5cm}|p{2.5cm}|X|}
    \hline
    \textbf{Notation} & \textbf{Description} & \textbf{Notation} & \textbf{Description} \\
    \hline
    $ \mathsf{did}$ & Decentralized Identifier &
    $ \mathsf{nonce}$ & A fresh nonce \\
    \hline
    $ \mathsf{genTxnPath}$ & Genesis transaction path &
    $ \mathsf{nymReq}$ & Public identity records with DID\\
    \hline   
    $ \mathsf{SW}$ & Steward wallet &
    $ \mathsf{req}$ & Request payload \\
    \hline 
    $ \mathsf{IL}$ & Indy Distributed Ledger & 
    $ \mathsf{W}$ & Write on ledger \\
    \hline
    $ \mathsf{R}$ & Read from ledger &
    $ \mathsf{connId}$ & Connection Id\\
    \hline   
    $ \mathsf{credDefId}$ & Credential Definition Id &
    $ \mathsf{credExId}$ & Credential Exchange Id\\
     \hline
    $ \mathsf{presExId}$ & Presentation Exchange Id &
    $ \mathsf{revRegId}$ & Revocation Registry Id \\
    \hline
  \end{tabularx}
  \label{tab:notations}
\end{table}

\subsection{Architecture of BDIMHS} 
\begin{figure}[!t]
    \centering
    \includegraphics[width=0.7\textwidth]{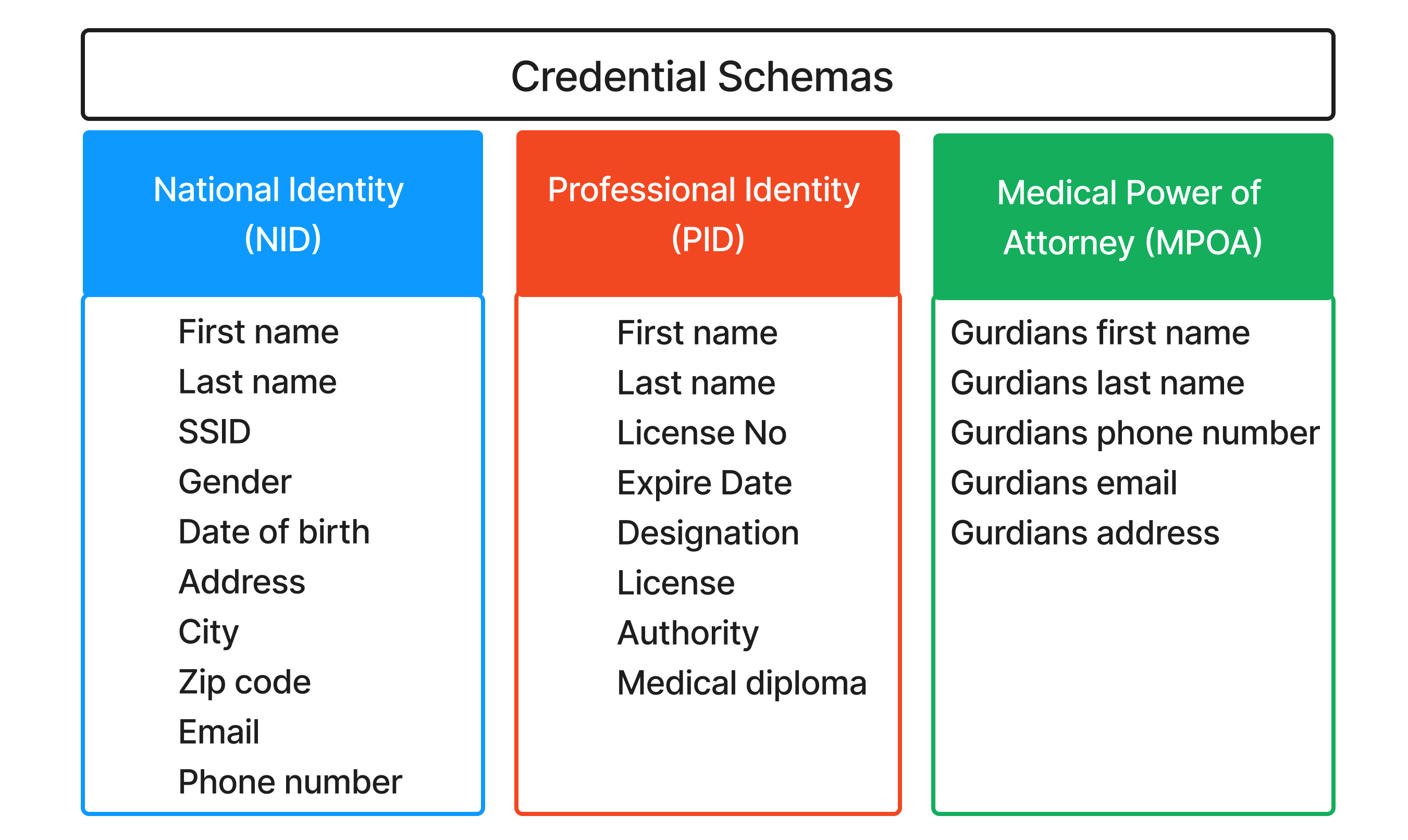} 
    \caption{Credential Schemas}
    \label{fig:credentialSchema}
\end{figure}

\begin{figure}[!t]
    \centering
    \includegraphics[width=\textwidth]{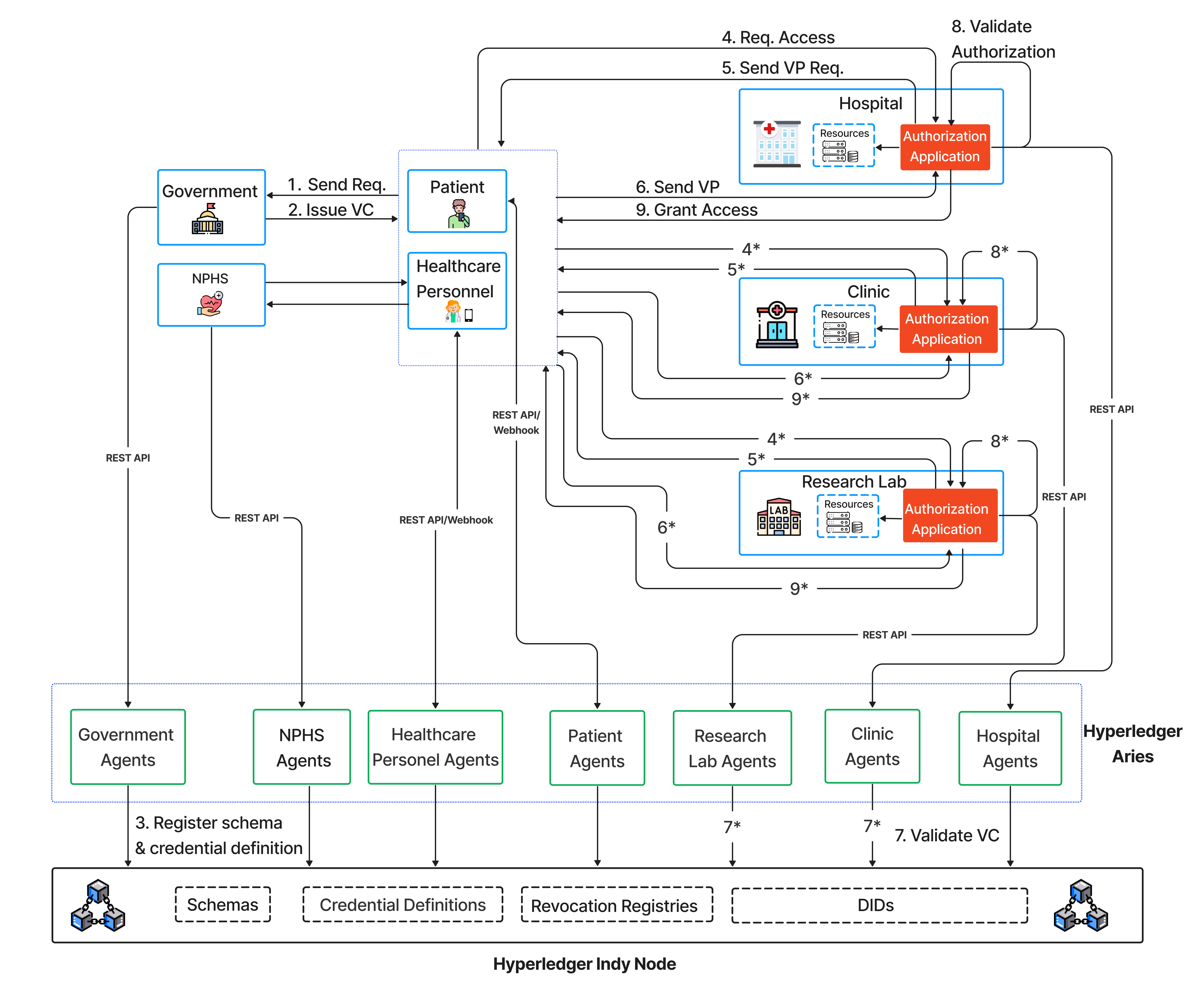} 
    \caption{High-level system architecture with multiple Verifiers}
    \label{fig:architectures}
\end{figure}

Hyperledger Indy furnishes the layer for decentralized identity by providing the Indy Node. Hyperledger Aries also provides the infrastructure for blockchain-rooted, peer-to-peer interactions and verifiable information exchange and maintains connectivity with the Indy Node. All the actors interact directly with Hyperledger Aries through their agents. Hyperledger Aries framework is a middleware between the client applications, the wallet and Indy framework. The system architecture of \scheme is illustrated in Figure~\ref{fig:architectures}.  In this scheme, we have initially defined three types of VCs that can be expanded later. One is National Identity (NID), Medical Power of Attorney (MPOA), and Professional Identity (PID). The first two credentials are common for all the \pt and \hp, and the latter is only for \hp. The NID provides the demographic data of a user. MPOA holds the power of attorney information in case of medical emergency and for minorities. The PID consists of professional data like healthcare personnel's license number, license expiry date, designation, medical diploma etc. Figure \ref{fig:credentialSchema} illustrates the credentials schemas used in the scheme.

Each \issuer, \holder and \verifier has their individual Aries agent as a form of web application or mobile application. \holder can establish a connection with both \issuer and \verifier to obtain VC or access to any service of the \verifier. \pt and \hp obtain their own VCs from \gov and \nphs. They can store the issued VC in their mobile wallet securely after accepting the issued VC. This stored VC can be used to prove the identity of the \pt or \hp when they try to access the service in \hosp or \cl or \rl. The \pt and \hp can select what attribute from the VC they need to share with the \verifier. Thus users maintain full control of their identity-related information. One of the essential components of this proposed scheme is the \authapp that all the \verifier uses. Here the \verifier acts as the OIDC client of our agent-based OIDC provider. Each \verifier has their own decentralized OIDC provider in the \authapp. This is due to avoiding the single point of failure of a single OIDC provider and stepping towards a decentralized OIDC provider. Upon successful verification of VC, the SSI OIDC Provider generates an ID token that contains the \holder's information and grants an access token based on the OAuth2.0 standard. The token is a JSON Web Token (JWT) containing information about the authenticated use with an expiration time and a nonce value to prevent replay attacks. 

\subsection{BDIMHS Workflow}
Workflow between the \pt, \hosp, and \gov have been considered here. For other actors, the workflow is the same. At first, the \gov creates a Schema Definition of PID based on the attributes defined above using the web application created for the \gov as \issuer. Then the \gov creates \emph{PID Credential Definition} based on PID schema. This Schema and Credential Definition are registered to the ledger and publicly available. Now the \gov is ready to issue credentials. 

\begin{figure}[!t]
    \centering
    \includegraphics[width=\textwidth]{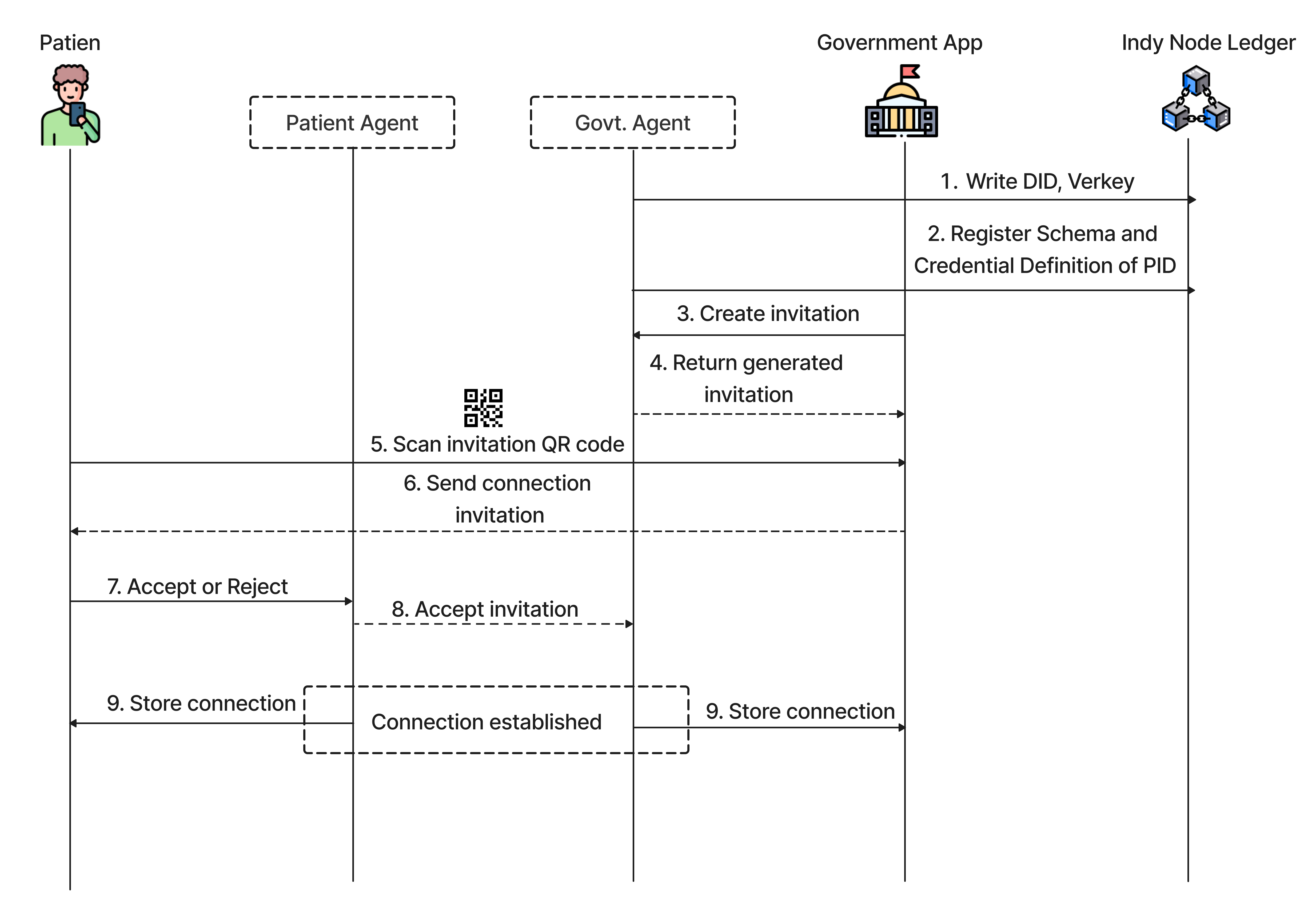} 
    \caption{Sequence diagram of registering DID, Schema, Credential Definition, and establishing Connection}
    \label{fig:sequence_connection}
\end{figure}

\noindent{\textbf{{Generate Invites and Start Connections: }}} 
The following steps are executed while creating invitations and connections. Figure~\ref{fig:sequence_connection} depicts the initialization and connection between \gov and \pt.

\begin{itemize}
    \item \gov generates an invitation.
    \item The invitation has been published in the \gov web application as a QR code. It can also be sent to the \pt personal email as an invitation URL.
    \item \pt scan the QR code in the Government web application.
    \item \pt\ either accepts or rejects the invitation request from the \gov.
    \item A pairwise DIDs has been generated after accepting the invitation to create a secure communication protocol.
    \item Both \gov and \pt now have a connection stored in their wallet with active status.
\end{itemize}

\begin{figure}[!t]
    \centering
    \includegraphics[width=\textwidth]{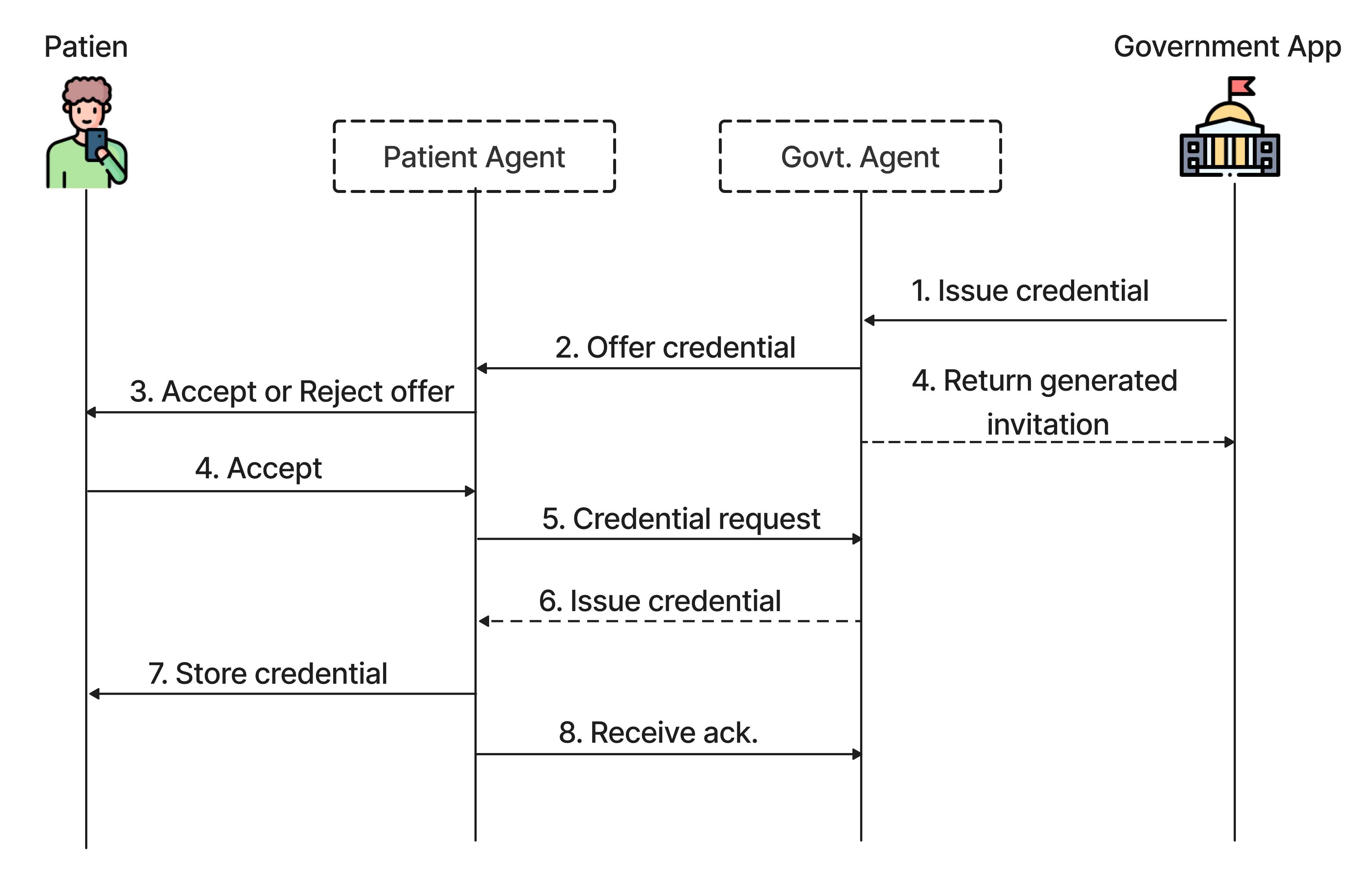} 
    \caption{Sequence diagram of credential issuance to Patient}
    \label{fig:sequence_issue}
\end{figure}

\noindent{\textbf{Government Issue Verifiable Credentials to Patient:}}
The following steps are performed while issuing credentials. 
Figure~\ref{fig:sequence_issue} depicts the VC issuance flow to the \pt.

\begin{itemize}
    \item \gov creates a VC based on PID and signs it with their private key.
    \item The \gov sends the credential offer to the patient using DIDComm.
    \item \pt gets the notification to accept and reject the offer. By accepting \pt sends the credential request to the \gov.
    \item \gov sent the VC to the \pt with the required attributes of PID along with metadata with \issuer DID, connection id, schema id, and credential definition id using DIDComm. 
    \item \pt accepts the VC and stores it in their wallet.
\end{itemize}

\begin{figure}[!t]
    \centering
    \includegraphics[width=\textwidth]{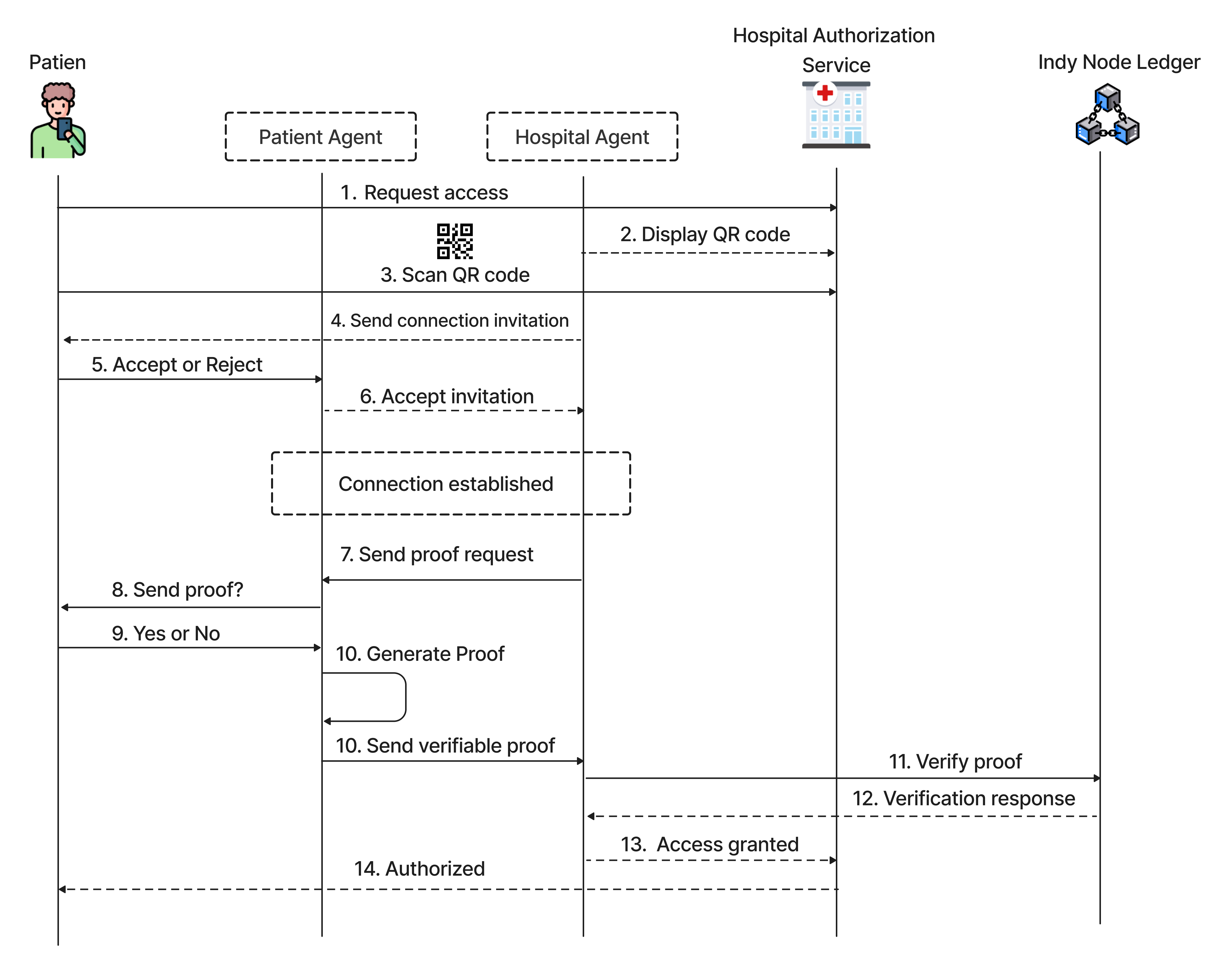} 
    \caption{Sequence diagram to verify proof and authorization to Hospital}
    \label{fig:sequence_proof}
\end{figure}

\noindent{\textbf{{Patient Authorization to Hospital:}}}
The following steps are performed while verifying the \pt. Figure~\ref{fig:sequence_proof} depicts the credential verification flow between \pt and \hosp.

\begin{itemize}
    \item \pt requests access to a set of resources in \hosp to access.
    \item The \hosp creates an invitation and displays it as a QR code.
    \item \pt scans the QR code from using the wallet application.
    \item \pt\ accepts the invitation request to establishes a secure connection.
    \item \authapp associated agent creates a proof request send it to the \pt. The proof request requires certain attributes that need to authenticate.
    \item \pt gets the notification of the proof request with the required attributes needed by the \hosp.
    \item \pt can either reject or accept and sends the proof with the required attributes from the stored VC to the \authapp.
    \item \authapp delegate the proof to the agent. The agent first validates the non-revocation by checking the \gov accumulator and then the authenticity of the credentials by checking the public key of the \gov from the ledger. Then, the VC has been verified in the \authapp.
    \item \pt grant access to the desired resources of the \hosp.
\end{itemize}

\section{Security Analysis}\label{sec:security_analysis}
We have evaluated the proposed scheme to provide individuals with secure and privacy-preserving healthcare authentication. We have examined various security aspects, including security and privacy requirements, threat modelling, consent management, and compliance with healthcare regulations. It is necessary to be noted that our proposed scheme does not address the security of internal servers or databases of the \verifier. It just provides the authentication and authorization mechanism layer for each internal resource.

\subsection{Threat Modelling}

To evaluate our proposed scheme, we have used the famous "STRIDE" model introduced by Microsoft. STRIDE model comprises six different threat categories, Spoofing, Tampering, Repudiation, Information disclosure, Denial of Service (DoS), and Elevation of privilege \cite{STRIDEMicrosoft}. 
Apart from the STRIDE, we have considered a few more threats that should be potential risks in the DIM system, Sybil attack, Replay attack, Identity Correlation, and Wallet attack.

\paragraph{\textbf{Spoofing}:}
In the field of DIMs, the first category Spoofing refers to the ability to perform illegal access to the \holder, \issuer or \verifier's identity actions. If malicious entities impersonate legitimate users of the device, they can access the stored verifiable credentials and associated private keys. Each wallet is primarily secured with biometric authentication. The \textit{Revocation} functionality prevents stealing VC and private keys. To avoid self-attestation claims, the \verifier will always check the validation of VC from the available DLT by validating the \issuer's DID. 

\paragraph{\textbf{Tampering}:}
On \issuer side, an adversary might change the configuration of credential issuance. On the other hand, \verifier can forge the inventory of trusted \issuer, which may lead to accepting VC created by intruder's Issuance authority \cite{GrunerMLM23}. The authenticity and reliability of the central issuance authorities exclude the possibility of the alteration of VC. Every issued VCs are digitally signed by \issuer's secret key (SK). The verifying authority will detect any forging attempt if the signature is invalid. 

\paragraph{\textbf{Repudiation}:}
Any user might purposefully do an unauthorized act that results in credential revocation and later denies doing the act and they can deliberately delete VCs from their wallet. Also, unlawful staff can revoke VC from \issuer's side. All the repudiation problems mentioned above can be monitored and prevented by the decentralized system with the tamper-proof decentralized ledger that keeps track of all the transactions happening on this identity network with timestamps.

\paragraph{\textbf{Information Disclosure}:}
Personal information can be disclosed while giving proof to the \verifier as well. Hyperledger Indy and Aries enforce strong cryptographic mechanisms to protect the confidentiality of VCs so that only the rightful owner can access them. Secure communication channel guarantees data integrity. This scheme also supports ZKP-based selective disclosure, ensuring minimal data disclosure. 

\paragraph{\textbf{Denial of Service}:}
The public endpoints exposed to connect to both \issuer or \verifier may be a potential target point of DOS attack, as these public endpoints are necessary to establish a secure connection. This can be mitigated with the implementation of Client puzzle \cite{AuraNL00} on the client application before creating connections. Nevertheless, the core decentralized network will always remain intact due to the nature of the public permissioned blockchain of Indy. 

\paragraph{\textbf{Elevation of Privilege}:}
\scheme will have a specific \issuer to issue VCs. The entitled \issuer will have their predefined data on whom to assign what. The VCs will hold certain information to identify which roles the \holder belongs to, and these claims can not be forged or tampered as we mentioned earlier. Also, the authorization server in the healthcare system will use the OIDC and OAuth2.0-based standard authentication and authorization mechanism that can significantly reduce this threat.  

\paragraph{\textbf{Sybil Attack}:}
In the permissioned blockchain, the Sybil attack is nearly impossible as only a limited number of pre-authorized participants in the Indy network known as \steward participate in the validation process. 

\paragraph{\textbf{Replay Attack}:}
It can be prevented by implanting secure communication protocols that provide message integrity and freshness. A random unique \texttt{id} property and \texttt{nonce} is required along with timestamps while using the DIDComm protocol that prevents replay attacks. On the \verifier side, the standard OIDC-based authentication prevents this by adding token expiration time and nonce inside JWT.

\paragraph{\textbf{Identity Correlation}:}
 One of the main goals of DIM is to prevent unauthorized access to personalized identity by introducing user-controlled VC, selective disclosure and ZKP. Hyperledger Indy is resistant to identity correlation.

\paragraph{\textbf{Wallet Attack}:}
Lost possession of the device that contains the identity wallet is one of the biggest threats. DKMS is used in the Hyperledger agents which make the wallet secure and provides a backup and recovery mechanism. Establishing DIDComm channels with malicious \verifier can expose Man-in-the-middle (MITM) attack \cite{KimCKKW21}. 

We assume the \holder as \pt and \hp will always act as honest users. Based on each category from the threat model, we have identified a total of 17 possible threat vectors \texttt{(T1--T17)} and a total of 15 countermeasures \texttt{(CM1--CM15)} in Table~\ref{table: STRIDE-theat-vector-counter-measure} that will mitigate these threats. By swiftly identifying and addressing these threats, the system maintains the integrity of the user identities healthcare system.

\begin{table}[!t]
\footnotesize
  \caption{Threat Model (I-Issuer, H-Holder, V-Verifier) }
  \label{table: STRIDE-theat-vector-counter-measure}
  \begin{tabularx}{\textwidth}{|c|X|X|}
    \hline
    \textbf{Category} & \textbf{Threat Vectors (T)} & \textbf{Countermeasures (CM)} \\
    \Xhline{3\arrayrulewidth}
    Spoofing & T1 - Impersonation as \holder (H) & CM1 - Biometric authentication \\
     & T2 - Stealing of VC and SK (H) & CM2 - Revocation functionality \\
     & T3 - Self-attested claims (H) & CM3- Registered Schema and Credential Definition \\
    \hline
    Tampering & T4 - Change VC (H) & CM4 - \issuer DID, signature on VC  \\
     & T5 - Trust unverified \issuer (V) & CM5 - Defined \issuer \\
    \hline
    Repudiation  & T6 - Revoke VC (I) & CM6 - Transaction timestamps \\
    \hline
    Information Disclosure & T7 - Eavesdropping (I, H, V) & CM7 - DIDComm \\
     & T8 - Credential exposure (H) & CM8 - ZKP \\
    \hline
    Denial of Service & T9 - Connection depletion (I, V)
     & CM9 - Permissioned Blockchain  \\ 
     & & CM10 - Client Puzzle\\
    \hline
    Elevation of Privilege & T10- Unauthorized privilege (V) & CM11 - RBAC  \\
     & T11 - Privilege escalation (V) & CM12 - VC based OIDC, CM2 \\
    \hline
    Sybil Attack & T12 - Fabrication of identities (H) & CM4 \\
    \hline
     Replay Attack & T13 - Capture DID while exchanging (I, H, V) & CM13 - DIDComm with timestamp and nonce \\
     & T14 - Reuse access token (V) & CM11  \\
    \hline
    Correlation Attack & T15 - Correlate credentials (H) & CM14 - Selective disclosure \\
    \hline
    Wallet Attack & T16 - Lost device (H) & CM15 - DKMS \\
     & T17 - MITM Attack (I, H, V) & CM8, CM13 \\
    \hline
  \end{tabularx}
\end{table}

\subsection{Security and Privacy Goals of DIM}
Confidentiality, integrity, and availability, commonly known as the CIA triad, are fundamental security goals in any system. However, in collaborative parameterized environments, the CIA triad alone is insufficient to address emerging threats and achieve comprehensive security. To overcome these limitations, an updated set of security goals has been proposed, including accountability, auditability, authenticity /trustworthiness, non-repudiation, and privacy \cite{CherdantsevaH13, Toorani16}. A comparison of the proposed scheme (\scheme) with other existing contemporary schemes, in terms of the security goals they provide, is presented in Table~\ref{table:requirements-comparison}.

\begin{table}[!t]
\centering
\caption{Comparison of security and privacy goals in different schemes}
\label{table:requirements-comparison}
\begin{tabularx}{\textwidth}{|p{0.18\linewidth}|Y|Y|Y|Y|Y|Y|Y|Y|Y|}
\hline
\textbf{Scheme} & \textbf{1} & \textbf{2} & \textbf{3} & \textbf{4} & \textbf{5} & \textbf{6} & \textbf{7} & \textbf{8}  \\
\Xhline{3\arrayrulewidth}
Health-ID \cite{JavedABMCQ21} & \ding{51} & \ding{51} & \ding{51} & \ding{51} & \ding{55} & \ding{51} & \ding{51} & \ding{55}  \\
\hline
BlockHIE \cite{JiangCWYMH18} & \ding{51} & \ding{51} & \ding{51} & \ding{51} & \ding{55} & \ding{51} & \ding{51} & \ding{51}  \\
\hline
MedRec \cite{AzariaEVL16} & \ding{51} & \ding{55} & \ding{55} & \ding{51} & \ding{51} & \ding{51} & \ding{51} & \ding{51}  \\
\hline
PBBIMUA \cite{XiangWF20} & \ding{51} & \ding{51} & \ding{51} & \ding{51} & \ding{55} & \ding{51} & \ding{51} & \ding{51}  \\
\hline
Manoj et al. \cite{ManojMN2022} & \ding{51} & \ding{51} & \ding{55} & \ding{51} & \ding{51} & \ding{51} & \ding{51} & \ding{51}  \\
\hline
HealthBlock \cite{ZaabarCJAA21} & \ding{51} & \ding{51} & \ding{51} & \ding{51} & \ding{55} & \ding{51} & \ding{51} & \ding{51} \\
\hline
DSMAC \cite{SaidiLAME22} & \ding{51} & \ding{51} & \ding{51} & \ding{51} & \ding{51} & \ding{51} & \ding{55} & \ding{51} \\
\hline
\textbf{BDIMHS}  & \ding{51} & \ding{51} & \ding{51} & \ding{51} & \ding{51} & \ding{51} & \ding{51} & \ding{51}  \\
\hline
\end{tabularx}
\begin{center}
1: Confidentiality \quad 2: Integrity \quad 3: Availability \quad 4: Accountability  \quad 5: Auditability \\ 6: Authenticity/Trustworthiness \quad 7: Non-repudiation \quad 8: Privacy 
\end{center}
\end{table}

\section{Performance Analysis}\label{sec:performance_analysis}
\begin{figure}[!htb]
\captionsetup{justification=centering}
\centering
\begin{minipage}{.48\textwidth}
  \centering
  \includegraphics[width=\textwidth]{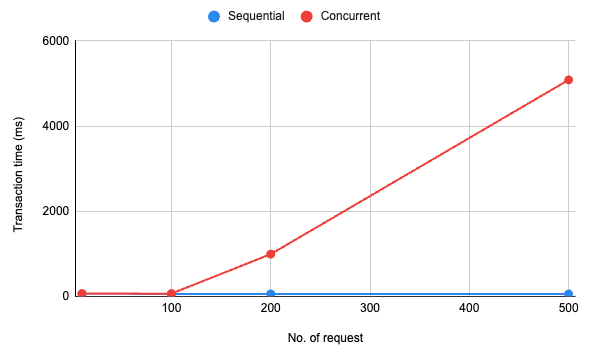}
    \caption{Connection invitation average transaction time}
    \label{fig:perf-connection}
\end{minipage}
\begin{minipage}{.48\textwidth}
  \centering
   \includegraphics[width=\textwidth]{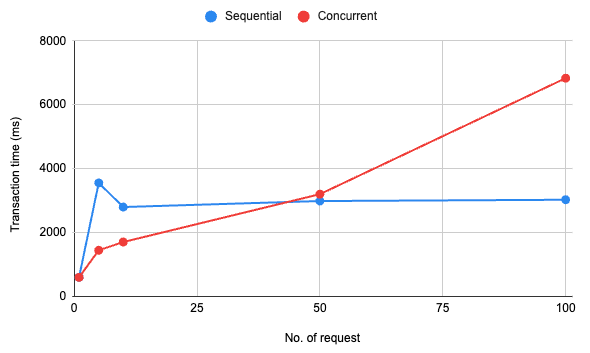}
    \caption{Register schema average transaction time}
    \label{fig:perf-schema}
\end{minipage}
\end{figure}

This section focuses on the performance analysis of the agents initiated as \issuer, \holder, and \verifier roles. For this analysis, we have considered the four nodes deployed using von-network. The performance analysis has been conducted on a MacBook Pro 13-inch with Apple M1 Silicon Chip 8-core CPU, 8-core GPU, 16GB unified Memory, 512GB SSD, and operating MacOS Ventura 13.3.1. We have used one docker container to run the nodes of von-network and separate containers for each agent. We have configured the resources for docker to a 4-core CPU, 8GB of memory and 1GB of swap memory. We have considered various performance metrics to assess the agent's capability to manage different workloads and performance, including transaction time, latency, throughput, standard deviation, resource utilization, bandwidth, and scalability.  
We have used Jmeter rampup period which determines how long it takes to "ramp up" to the number of threads/users chosen. We can see that sequential execution of the connection invitation maintains a consistent performance with low transaction time, minimal standard deviation and stable throughput. Nevertheless, for concurrent requests, when there is a large number of requests simultaneously, the transaction time increases with the number of user, illustrated in Table~\ref{table:MergedConnection}. The standard deviation also fluctuates a lot with many concurrent users. Figure~\ref{fig:perf-connection} depicts the visualization of the average transaction time of connection invitation.

\begin{table}[!t]
  \centering
  \caption{Connection Invitation for Rampup=1 and sequential requests. (* Values in parentheses correspond experiments for Rampup=10 and concurrent requests.)}
  \begin{tabularx}{\textwidth}
  {|p{1.8cm}|p{2cm}|p{2.2cm}|p{2.6cm}|p{2.6cm}|X|}
    \hline
    \textbf{\# Requests} & \textbf{Min (ms)} & \textbf{Max (ms)} & \textbf{Transaction Time Avg. (ms)} & \textbf{Standard Deviation} & \textbf{Throughput} \\
    \Xhline{3\arrayrulewidth}
    10 & 52 \rr{(58*)} & 70 \rr{(88*)} & 61 \rr{(67*)} & 5.09 \rr{(1.9*)} & 16.2 \rr{(1.1*)} \\
    \hline
    100 & 51 \rr{(56*)} & 82 \rr{(99*)} & 60 \rr{(66*)} & 5.65 \rr{(8.1*)} & 16.1 \rr{(10*)} \\
    \hline
    200 & 51 \rr{(223*)} & 121 \rr{(1767*)} & 59 \rr{(993*)} & 6.91 \rr{(43.8*)} & 16.5 \rr{(18*)} \\
    \hline
    500 & 50 \rr{(258*)} & 250 \rr{(7823*)} & 59 \rr{(5085*)} & 6.95 \rr{(202.5*)} & 16.4 \rr{(17.3*)}\\
    \hline
  \end{tabularx}
  \label{table:MergedConnection}
\end{table}

\begin{table}[!t]
  \centering
  \caption{Register Schema for Rampup=1 and sequential requests. (*Values in parentheses correspond experiments for Rampup=10 and concurrent requests.)}
  \begin{tabularx}{\textwidth}{|p{1.8cm}|p{2cm}|p{2.2cm}|p{2.6cm}|p{2.6cm}|X|}
    \hline
    \textbf{\# Requests} & \textbf{Min (ms)} & \textbf{Max (ms)} & \textbf{Transaction Time Avg. (ms)} & \textbf{Standard Deviation} & \textbf{Throughput} \\
    \Xhline{3\arrayrulewidth}
    1 & 599 \rr{(599*)} & 599 \rr{(599*)} & 599 \rr{(599*)} & 0 \rr{(0*)} & 1.7 \rr{(1.7*)}\\
    \hline  
    5 & 599 \rr{(580*)} & 3052 \rr{(3687*)} & 3554 \rr{(1446*)} & 972.6 \rr{(76.3*)} & 0.38 \rr{(0.5*)} \\
    \hline    
    10 & 691 \rr{(618*)} & 3077 \rr{(2888*)} & 2799 \rr{(1704*)} & 703.1 \rr{(85.3*)} & 0.35 \rr{(1*)} \\
    \hline
    50 & 603 \rr{(790*)} & 3082 \rr{(4873*)} & 2987 \rr{(3202*)} & 341.63 \rr{(95.6*)} & 0.35 \rr{(4.4*)} \\
    \hline
    100 & 605 \rr{(864*)} & 4449 \rr{(9138*)} & 3027 \rr{(6833*)} & 284.57 \rr{(135.9*)} & 0.3 \rr{(6*)} \\
    \hline
  \end{tabularx}
  \label{table:RegisterSchemaMerged}
\end{table}

\begin{figure}[!t]
\captionsetup{justification=centering}
\centering
\begin{minipage}{.5\textwidth}
  \centering
  \includegraphics[width=\textwidth]{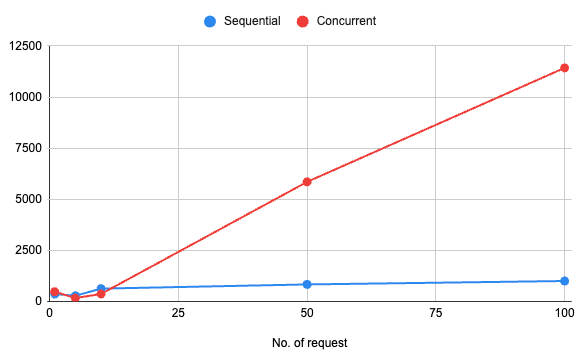}
    \caption{Issue credential average transaction time}
    \label{fig:perf-issue}
\end{minipage}
\begin{minipage}{.5\textwidth}
  \centering
   \includegraphics[width=\textwidth]{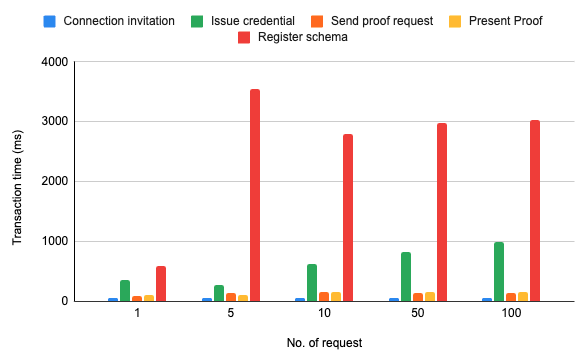}
    \caption{Transaction time comparison of the functions}
    \label{fig:perf-compare-all}
\end{minipage}
\end{figure}

The registration of schema definition involves writing operations on the ledger. As a result, for sequential and concurrent operations, the transaction time is moderate based on the number of requests illustrated in Table~\ref{table:RegisterSchemaMerged}. Figure~\ref{fig:perf-schema} depicts the visualization of the average transaction time of registering schema. Issue credential maintains a stable transaction time for sequential requests, an average time of 1000ms for 100 requests. However, for 100 concurrent requests, it takes a high time of 11425ms shown in Table~\ref{table:IssueMerged} and in Figure~\ref{fig:perf-issue}. 

\begin{table}[!t]
  \centering
  \caption{Issue Credential for Rampup=1 and sequential requests. (*Values in parentheses correspond experiments for Rampup= 10 and concurrent requests.)}
  \begin{tabularx}{\textwidth}{|p{1.8cm}|p{2cm}|p{2.5cm}|p{2.5cm}|p{2.5cm}|X|}
    \hline
    \textbf{\# Requests} & \textbf{Min (ms)} & \textbf{Max (ms)} & \textbf{Transaction Time Avg. (ms)} & \textbf{Standard Deviation} & \textbf{Throughput} \\
    \Xhline{3\arrayrulewidth}
    1 & 362 \rr{(367*)} & 362 \rr{(711*)} & 362 \rr{(475*)} & 0 \rr{(0*)} & 2.8 \rr{(2.1*)} \\
    \hline  
    5 & 182 \rr{(120*)} & 350 \rr{(291*)} & 278 \rr{(163*)} & 5.8 \rr{(6*)} & 3.6 \rr{(0.6*)} \\
    \hline   
    10 & 204 \rr{(157*)} & 1557 \rr{(711*)} & 628 \rr{(367*)} & 51.74 \rr{(17.7*)} & 1.6 \rr{(1*)} \\
    \hline    
    50 & 230 \rr{(320*)} & 2086 \rr{(11820*)} & 832 \rr{(5850*)} & 45.17 \rr{(379.1*)} & 1.2 \rr{(2.3*)} \\
    \hline
    100 & 192 \rr{(364*)} & 3810 \rr{(26437*)} & 1000 \rr{(11425*)} & 72.43 \rr{(779.7*)} & 1 \rr{(2.8*)} \\
    \hline
  \end{tabularx}
  \label{table:IssueMerged}
\end{table}

The performance of sending proof request operation is fast for both sequential and concurrent requests. It maintains low transaction time and high throughput discarding numerous requests depicts in Table~\ref{table:sendProofMerged}. Furthermore, the present proof maintains a steady transaction time, standard deviation, and throughput despite the number of requests illustrated in Table~\ref{table:Perf-present-proof}.
Another evaluation has been performed with initiating two agents in the network and simulated the startup to credential exchange scenario sequentially with 10, 100 and 300 credential exchanges; the results are demonstrated in Table~\ref{tab:process-time}. This evaluation proves that the system is highly scalable. 

\begin{table}[!t]
  \centering
  \caption{Send Proof Request for Rampup=1 and sequential requests. (*Values in parentheses correspond experiments for Rampup= 10 and concurrent requests.)}
  \begin{tabularx}{\textwidth}{|p{1.8cm}|p{2cm}|p{2.5cm}|p{2.5cm}|p{2.5cm}|X|}
    \hline
    \textbf{\# Request} & \textbf{Min (ms)} & \textbf{Max (ms)} & \textbf{Transaction Time Avg. (ms)} & \textbf{Standard Deviation} & \textbf{Throughput} \\
    \Xhline{3\arrayrulewidth}
    1 & 99 \rr{(99*)} & 99 \rr{(99*)} & 99 \rr{(99*)} & 0 \rr{(0*)} & 10.1 \rr{(10.1*)}\\
    \hline  
    5 & 74 \rr{(68*)} & 164 \rr{(86*)} & 134 \rr{(73*)} & 33.05 \rr{(6.5*)} & 7.4 \rr{(0.62*)}\\
    \hline    
    10 & 79 \rr{(67*)} & 252 \rr{(131*)} & 156 \rr{(75*)} & 45.37 \rr{(18.61*)} & 6.4 \rr{(1.1*)}\\
    \hline
     50 & 88 \rr{(78*)} & 250 \rr{(162*)} & 148 \rr{(118*)} & 35.01 \rr{(22.9*)} & 6.7 \rr{(5*)}\\
    \hline
    100 & 85 \rr{(92*)} & 195 \rr{(4377*)} & 139 \rr{(2320*)} & 24.46 \rr{(136.4*)} & 7.1 \rr{(7.1*)}\\
    \hline
  \end{tabularx}
  \label{table:sendProofMerged}
\end{table}

\begin{table}[!t]
  \centering
  \caption{Present Proof; where Rampup is 1 and request is Sequential}
  \begin{tabularx}{\textwidth}{|p{1.7cm}|X|X|p{2.5cm}|p{1.8cm}|X|}
    \hline
    \textbf{\# Request} & \textbf{Min (ms)} & \textbf{Max (ms)} & \textbf{Transaction Time Avg. (ms)} & \textbf{Standard Deviation} & \textbf{Throughput} \\
    \Xhline{3\arrayrulewidth}
    1 & 111 & 111 & 111 & 0 & 8.9\\
    \hline  
    5 & 80 & 130 & 110 & 19.35 & 8.9\\
    \hline    
    10 & 112 & 208 & 161 & 31.92 & 6.2\\
    \hline
     50 & 79 & 282 & 160 & 36.21 & 6.2\\
    \hline
    100 & 101 & 263 & 163 & 28.44 & 6.1\\
    \hline
  \end{tabularx}
  \label{table:Perf-present-proof}
\end{table}

\begin{table}[!t]
  \centering
  \caption{Time duration for 10, 100, and 300 credential exchange}
  \begin{tabularx}{\textwidth}{|X|X|X|X|}
    \hline
    \textbf{Process} & \textbf{Time (s) for 10} & \textbf{Time (s) for 100 } & \textbf{Time (s) for 300}  \\
    \hline
    Startup & 32.9 & 32.92  & 30.66 \\
    \hline
    Connection & 0.85 & 0.89 & 0.91 \\
    \hline
    Register Schema & 4.12 & 8.07 & 9.91 \\
    \hline
    Exchange Credential & 10.21 & 83.56 & 217.1 \\
    \hline
  \end{tabularx}
  \label{tab:process-time}
\end{table}

The core functionalities maintain optimal performance in sequential requests and can also tolerate large concurrent requests. The transaction time comparison of all the functionalities depicts in Figure~\ref{fig:perf-compare-all}. However, we can see that the standard deviation of all the operations deviates significantly, which echoes the performance degradation. The degradation could be the reason for the computational capability of the server, as we are using containerized docker as the server. 

\begin{figure}[!t]
\captionsetup{justification=centering}
\centering
\begin{minipage}{.5\textwidth}
    \centering
    \includegraphics[width=\textwidth]{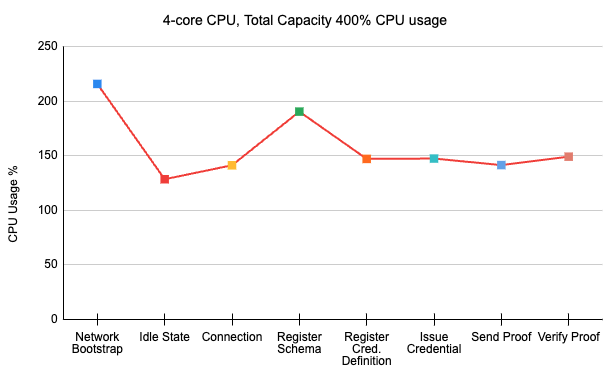}
    \caption{CPU usage of primary processes}
    \label{fig:cpu}
\end{minipage}%
\begin{minipage}{.5\textwidth}
  \centering
    \includegraphics[width=\textwidth]{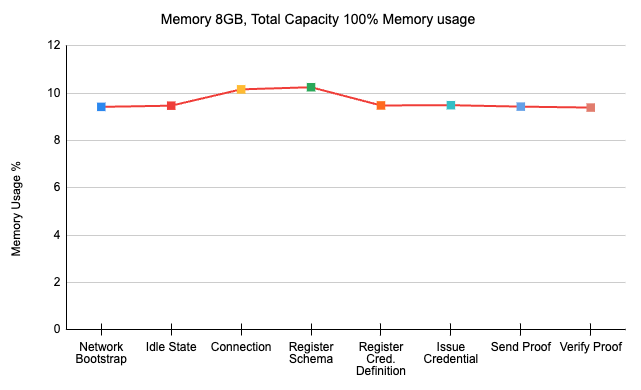}
    \caption{Memory usage of primary processes}
    \label{fig:memory}
\end{minipage}
\end{figure}

\paragraph{\textbf{Resource Utilization:}} We have conducted the resource utilization in a specific scenario where we have performed all the functionality required to finish a complete identity management flow, from bootstrapping the network to issuing credentials and verifying the proof, which involves \gov, \pt, and \hosp agents. Most of the required functionalities are executed once to measure resource utilization. Throughout the process, CPU usage remains optimal for all the read operations. The visualization of CPU and memory usage is depicted in Figure~\ref{fig:cpu} and Figure~\ref{fig:memory} correspondingly. Overall, the proposed scheme shows significant efficiency in terms of resource utilization.

\paragraph{\textbf{Interoperability:}}
The proposed scheme enforces interoperability by implementing a standard identity management system that facilitates the seamless exchange and integration of authentication across different healthcare entities. \scheme leverages widely adopted identity W3C standards VC and DIDs, providing a common framework for issuing and verifying digital identity credentials across healthcare systems. \scheme promotes using of standardized digital credentials by utilizing standard Credential Schema and Credential Definition. The \issuer publishes a standard set of Credential Definition which anyone on this network can use, and it ensures standardization of the digital credentials.

\section{Conclusion}\label{sec:conclusion}
In this paper, we proposed \scheme, a passwordless, secure, and user-centric identity management scheme that addresses the limitations and challenges of traditional identity management. By leveraging the permissioned blockchain feature of Hyperledger Indy and Hyperledger Aries, \scheme ensures regulatory compliance without storing personal identity-related data on the blockchain, as every credential remains in the possession of the user. 
Through the integration of w3c standard VC and DID, \scheme establishes a decentralized trust triangle between the identity provider, user, and verifier, creating a crucial trust layer. To demonstrate the feasibility of \scheme, we developed a PoC tailored for the healthcare system, catering to the needs of issuers, holders, and verifiers. Furthermore, we conducted a comprehensive security analysis using the STRIDE threat model, comparing \scheme with contemporary alternatives in the same domain. Strategies and countermeasures to address challenges and achieve security and privacy goals when implementing a DIM in the healthcare system were also explored.
To assess the system's performance, we evaluated scalability, efficiency, and resource utilization, demonstrating the efficiency, scalability, and potential for system interoperability provided by \scheme. By presenting the architecture, PoC, implementation details, and conducting comprehensive security and performance analyses, we have demonstrated the relevance and significance of \scheme for healthcare applications.

\section*{Acknowledgement}
This work was partially supported by the Norwegian Research Council under project number 331903.

\bibliographystyle{IEEEtran}
\small{\bibliography{Bib}}

\end{document}